\documentclass[paper,notoc]{JHEP3}

\usepackage{epsfig}
\usepackage{graphicx}
\usepackage{feynmp}

\usepackage{amsmath}
\usepackage{amsfonts}
\usepackage{amssymb}
\usepackage{graphicx}

\def\beq{\begin{equation}}
\def\eeq{\end{equation}}
\newcommand{\bea}{\begin{eqnarray}}
\newcommand{\eea}{\end{eqnarray}}
\newcommand{\nn}{\nonumber}

\def\Eqn#1{Eq.~(\ref{#1})}

\def\eqns#1#2{Eqs.~(\ref{#1}) and~(\ref{#2})}

\def\sec#1{Section~{\ref{#1}}}

\def\bsp#1\esp{\begin{split}#1\end{split}}

\newcommand{\eps}{\epsilon}

\newcommand{\ord}{\begin{cal}O\end{cal}}

\newcommand{\cC}{{\cal C}}

\def\P{{\cal P} }

\def\N{{\cal N} }
\def\C{{\cal C} }

\newcommand{\rd}{\mathrm{d}}

\def\bit#1\eit{\begin{itemize}#1\end{itemize}}
\def\ben#1\een{\begin{enumerate}#1\end{enumerate}}

\newenvironment{sloppyequation}[0]{\sloppy\begin{flushleft}\hspace*{0.75cm}\(}{\)\end{flushleft}\fussy}
\newenvironment{sloppytext}[0]{\sloppy\begin{flushleft}}{\end{flushleft}\fussy}

\newcommand{\beqsloppy}{\begin{sloppyequation}}
\newcommand{\eeqsloppy}{\end{sloppyequation}}
\newcommand{\btxtsloppy}{\begin{sloppytext}}
\newcommand{\etxtsloppy}{\end{sloppytext}}

\title{A Two-Loop Octagon Wilson Loop in $\N = 4$ SYM}

\author{Vittorio Del Duca\\
PH Department, TH Unit, CERN CH-1211, Geneva 23, Switzerland\\
INFN, Laboratori Nazionali Frascati, 00044 Frascati (Roma), Italy\\
       E-mail: \email{vittorio.del.duca@cern.ch}}

\author{Claude Duhr\\
Institute for Particle Physics Phenomenology,
University of Durham\\ Durham, DH1 3LE, U.K.\\
E-mail: \email{claude.duhr@durham.ac.uk}}

\author{Vladimir A. Smirnov\\
Nuclear Physics Institute of Moscow State University\\
Moscow 119992, Russia\\
E-mail: \email{smirnov@theory.sinp.msu.ru}}

\abstract{
In the planar $\begin{cal}N\end{cal}=4$ supersymmetric Yang-Mills theory at weak coupling,
we perform the first analytic computation of a two-loop eight-edged Wilson loop embedded into the boundary of $AdS_3$.
Its remainder function is given as a function of uniform transcendental weight four in terms of
a constant plus a product of four logarithms. We compare to the strong-coupling result,
and test a conjecture on the universality of the remainder function proposed in the literature.}

\keywords{N=4 SYM}
\preprint{IPPP/10/47, DCPT/10/94\\ CERN-PH-TH/2010-141}

\begin{document}


\section{Introduction}
\label{sec:intro}

In the planar $\begin{cal}N\end{cal}=4$ supersymmetric Yang-Mills (SYM) theory at strong coupling,
Alday and Maldacena~\cite{Alday:2007hr} noted a duality between an $n$-point colour-stripped
scattering amplitude and the vacuum expectation value of a Wilson loop with a contour made
by $n$ light-like edges, whose length and ordering is given by the momentum and ordering
of the particles in the amplitude. Remarkably, such a duality was found to hold also
for the planar $\begin{cal}N\end{cal}=4$ SYM theory at weak coupling,
between the one-loop four-point amplitude in a maximally-helicity violating (MHV)
configuration and a one-loop four-edged Wilson loop
with the contour defined as above~\cite{Drummond:2007aua}.

At one-loop level, the duality has been extended to Wilson loops
and MHV amplitudes with an arbitrary number of points~\cite{Brandhuber:2007yx}.
Writing at any loop order $L$ the amplitude $M_n^{(L)}$ as the tree-level amplitude,
$M_n^{(0)}$, which depends on the helicity configuration, times a scalar function,
$m_n^{(L)}$, and introducing a Wilson-loop coefficient, $w_n^{(L)}$,
the duality is expressed  by $w_n^{(1)} = m_n^{(1)} + \textrm{const} +\ord(\epsilon)$.
At two-loop level, the duality has been successfully tested on the four-edged Wilson loop and the
four-point MHV amplitude~\cite{Drummond:2007cf}, and on the
five-edged~\cite{Drummond:2007au} and
six-edged~\cite{Drummond:2007bm,Drummond:2008aq} Wilson loops
and the parity-even part of the five-point~\cite{Bern:2006vw} and six-point~\cite{Bern:2008ap}
MHV amplitudes.

The $L$-loop Wilson loop fulfills a Ward identity for a special conformal boost,
whose solution for $L\ge 2$ can be written as the sum of two contributions: a term
which iterates the structure of the one-loop Wilson loop, augmented,
for $n\ge 6$, by a function $R_{n,WL}^{(L)}$ of conformally
invariant cross ratios~\cite{Drummond:2007au}. The iterative term is known through
a conjecture on the structure of
the MHV amplitudes~\cite{Anastasiou:2003kj,Bern:2005iz}, while $R_{n,WL}^{(L)}$,
which is termed the \emph{remainder function}, is not fixed by the Ward identity
and must be computed. Writing likewise the $L$-loop MHV amplitude as the sum
of the iterative formula of the one-loop amplitude plus a function $R_{n}^{(L)}$
of conformally invariant cross ratios, and choosing appropriately the relative normalisation
of one-loop amplitudes and Wilson loops, the duality between MHV amplitudes and Wilson loops
is then stated by the equality of their remainder functions, $R_{n}^{(L)} = R_{n,WL}^{(L)}$.
At two loops, $R_{n}^{(2)}$ is known numerically for $n=6$~\cite{Bern:2008ap},
while $R_{n,WL}^{(2)}$ is known for arbitrary $n$ through a numerical
algorithm~\cite{Anastasiou:2009kna}, which has been used to compute it for up to
30 edges~\cite{Brandhuber:2009da}. The two-loop six-edged remainder function
$R_{6,WL}^{(2)}$ is also known analytically as a function of uniform transcendental
weight four in terms of multiple polylogarithms in three conformally
invariant cross ratios~\cite{DelDuca:2009au,DelDuca:2010zg}\footnote{In Ref.~\cite{Zhang:2010tr},
$R_{6,WL}^{(2)}$ has been expressed in terms of one-dimensional integrals.}.
Beyond two loops, the remainder functions are unknown.

At strong coupling, the six-edged remainder function has been computed
analytically in the limit where all three cross ratios are equal~\cite{Alday:2009dv}. In contrast to the weak coupling result,
which is of uniform and intrinsic transcendental weight four~\cite{DelDuca:2010zg} (intrinsic
in the sense that the terms of the polynomial of weight four cannot usually be reduced to the product
of terms of lower weight), the strong coupling result is expressed as a combination of terms with
different transcendentality. Although analytically the remainder functions at strong
and weak coupling are different functions, it is worth noting that they are numerically
very close over a wide range of values for the conformal ratio, albeit with significative
differences~\cite{DelDuca:2010zg, Alday:2009dv}.

At strong coupling, for Wilson loops embedded into the boundary of $AdS_3$,
the eight-edged remainder function has been computed in terms of two variables, $\chi^+$ and
$\chi^-$~\cite{Alday:2009yn}. The strong-coupling result has been compared numerically to the weak-coupling
two-loop result~\cite{Brandhuber:2009da}. In this paper, we present the first analytic calculation of
the two-loop eight-edged Wilson loop as a function of $\chi^+$ and $\chi^-$. The ensuing
remainder function takes a remarkably simple form and is given, up to a constant, by the product of four logarithms.

Our paper is organized as follows: In \sec{sec:wl} we introduce our notation and conventions, and we review the eight-edged Wilson loop both in general as well as in the special kinematics of Ref.~\cite{Alday:2009yn}. In Section~\ref{sec:R8} we discuss our strategy for the computation of the octagon remainder function and we present the main result of this paper, the analytic expression for $R_{8,WL}^{(2)}(\chi^+,\chi^-)$. 
In order to test the conjectures made in Ref.~\cite{Brandhuber:2009da}, in Section~\ref{sec:strong} we compare our result to the strong-coupling result of Ref.~\cite{Alday:2009yn}. In Section~\ref{sec:conclusion} we draw our conclusions.

\section{The two-loop Wilson loop}
\label{sec:wl}

\subsection{Definitions}

The Wilson loop is defined through the path-ordered exponential,
\beq
W[\C_n] = {\rm Tr}\ \P\ {\rm exp} \left[ ig \oint {\rm d}\tau \dot{x}^\mu(\tau) A_\mu(x(\tau)) \right]\,,
\label{eq:wloop}
\eeq
computed on a closed contour $\C_n$. In what follows, the closed contour
is a light-like $n$-edged polygonal contour~\cite{Alday:2007hr}.
The contour is such that labelling the $n$ vertices of the polygon as $x_1,\ldots,x_n$, the distance
between any two contiguous vertices, {\em i.e.}, the length of the edge in between, is given by the
momentum of a particle in the corresponding colour-ordered scattering amplitude,
\beq
p_i = x_i - x_{i+1}\, ,\label{eq:dist}
\eeq
with $i=1,\ldots,n$. Because the $n$ momenta add up to zero, $\sum_{i=1}^n p_i = 0$, the $n$-edged contour
closes, provided we make the identification $x_1 = x_{n+1}$.

In the weak-coupling limit, the Wilson loop can be computed as an expansion in the coupling.
The expansion of \Eqn{eq:wloop} is done through
the non-abelian exponentiation theorem~\cite{Gatheral:1983cz,Frenkel:1984pz},
which gives the vacuum expectation value of the Wilson loop as an exponential,
\beq
\langle W[\C_n] \rangle = 1 + \sum_{L=1}^\infty a^L W_n^{(L)} = {\rm exp} \sum_{L=1}^\infty a^L w_n^{(L)}\, ,\label{eq:fgt}
\eeq
where the coupling is defined as
\beq
a = \frac{g^2 N}{8\pi^2}\,.
\eeq
For the first two loop orders, one obtains
\beq
w_n^{(1)} = W_n^{(1)}\,, \qquad
w_n^{(2)} = W_n^{(2)} - \frac{1}{2} \left( W_n^{(1)}\right)^2\, .\label{eq:twowl}
\eeq
The one-loop coefficient $w_n^{(1)}$ was evaluated in Refs.~\cite{Drummond:2007aua,Brandhuber:2007yx},
where it was given in terms of the one-loop $n$-point MHV amplitude,
\beq
w_n^{(1)} = \frac{\Gamma(1-2\eps)}{\Gamma^2(1-\eps)} m_n^{(1)} =
m_n^{(1)} - n \frac{\zeta_2}{2} + \ord(\eps)
\, ,\label{eq:wlamp}
\eeq
with $\zeta_i = \zeta(i)$ and $\zeta(z)$ the Riemann zeta function, and
where the amplitude is a sum of one-loop {\em two-mass-easy} box functions~\cite{Bern:1994zx},
\beq
m_n^{(1)} = \sum_{p,q} F^{\rm 2m e}(p,q,P,Q)\,, \label{eq:2me}
\eeq
where $p$ and $q$ are two external momenta corresponding to two opposite massless legs,
while the two remaining legs $P$ and $Q$ are massive. The two-loop coefficient $w_n^{(2)}$
has been computed analytically for $n=4$~\cite{Drummond:2007cf}, $n=5$~\cite{Drummond:2007au} and
$n=6$~\cite{DelDuca:2009au,DelDuca:2010zg}, and
numerically for $n=6$~\cite{Drummond:2008aq} and $n=7, 8$~\cite{Anastasiou:2009kna}.

The Wilson loop fulfils a special conformal Ward identity~\cite{Drummond:2007au},
whose solution is an iterative formula over the
one-loop Wilson loop~\cite{Anastasiou:2003kj,Bern:2005iz} plus, for $n\ge 6$, an arbitrary function
of the conformally invariant cross ratios. Thus, the two-loop coefficient $w_n^{(2)}$ can be written as
\beq
w_n^{(2)}(\eps) = f^{(2)}_{WL}(\eps)\, w_n^{(1)}(2\eps) + C_{WL}^{(2)} + R_{n,WL}^{(2)}
+ \ord(\eps)\, ,\label{eq:wl2wi}
\eeq
where the constant is $C_{WL}^{(2)} = -\zeta_2^2/2$,
and the function $f^{(2)}_{WL}(\eps)$ is~\cite{Drummond:2007cf,Anastasiou:2009kna,Korchemskaya:1992je},
\beq
f^{(2)}_{WL}(\eps) = -\zeta_2 + 7\zeta_3\eps - 5\zeta_4\eps^2\, .\label{eq:fdue}
\eeq
With the two-loop coefficient $w_n^{(2)}$ given by \eqns{eq:wl2wi}{eq:fdue} and a similar expansion
of the two-loop MHV amplitude~\cite{Anastasiou:2003kj},
\beq
m_n^{(2)}(\eps) = \frac{1}{2} \left[m_n^{(1)}(\eps)\right]^2
+ f^{(2)}(\eps)\, m_n^{(1)}(2\eps) + C^{(2)} + R_{n}^{(2)} + \ord(\eps)\, ,\label{eq:ite2bds}
\eeq
with the constant $C^{(2)} = C_{WL}^{(2)}$, and the function $f^{(2)}(\eps) = -\zeta_2 - \zeta_3\eps - \zeta_4\eps^2$,
the duality between Wilson loops and amplitudes
is expressed by the equality of their remainder functions~\cite{Anastasiou:2009kna},
\beq
R_{n,WL}^{(2)} = R_n^{(2)}\, .
\eeq

\subsection{The two-loop eight-edged Wilson loop}
\label{sec:2loopwl}

The diagrams that enter the computation of the two-loop eight-edged Wilson loop have been given explicitly in
Ref.~\cite{Anastasiou:2009kna} in terms of multifold Feynman parameter-like integrals. In total, 49 different diagrams (plus their cyclic permutations over the external edges) contribute to $w^{(2)}_8$. The most difficult contribution comes from the so-called ``hard diagram", which corresponds to the situation where three different edges of the polygon are connected by a three-gluon vertex. However, the complexity of the problem does not only come from the large number of diagrams, but also from the fact that the Wilson loop is a function of twenty cyclic Mandelstam invariants\footnote{We do not impose the Gram determinant constraint, {\it i.e.}, we no not restrict the external momenta to four
dimensions.} of the form $s_{i,i+1}$, $s_{i,i+1,i+2}$ and $s_{i,i+1,i+2,i+3}$ (See Appendix~\ref{app:kinematics} for a summary of the eight-point kinematics).

From the special conformal Ward identity~\cite{Drummond:2007au}, we expect that the two-loop eight-edged
remainder function $R_{8,WL}^{(2)}$, defined through \Eqn{eq:wl2wi}, be a function of cross ratios only,
defined as,
\beq
u_{ij} = \frac{x_{ij+1}^2 x_{i+1j}^2}{x_{ij}^2x_{i+1j+1}^2}\,,
\eeq
where $x_{ij}^2 = (x_i-x_j)^2= s_{i,i+1,\ldots,j-1}$, with $j > i$.
For the eight-edged Wilson loop there are twelve cross ratios, which are given explicitly in \Eqn{eq:8cross} and which
can be divided into two groups of eight and four respectively,
\beq\label{eq:us8}
u_{i,i+3} = {x_{i,i+4}^2\,x_{i+1,i+3}^2\over x_{i,i+3}^2\,x_{i+1,i+4}^2} \qquad {\rm and} \qquad u_{k,k+4} = {x_{k,k+5}^2\,x_{k+1,k+4}^2\over 
x_{k,k+4}^2\,x_{k+1,k+5}^2}\,,
\eeq
with $i=1,\ldots,8$ and $k=1,\ldots,4$.
Hence in general kinematics the eight-point remainder function $R_{8,WL}^{(2)}$ is a function of twelve variables. Currently that lies beyond our technical capabilities. Therefore it is natural to look for special kinematics where the twelve cross ratios are parametrised by a smaller set of free parameters.
In Ref.~\cite{Alday:2009yn}, the eight-point remainder function was computed at strong coupling for a Wilson loop embedded into the boundary of $AdS_3$.  In this set-up all twelve conformal ratios can be expressed in terms of two real positive parameters $\chi^\pm$. Explicitly, the relations read,
\beq\bsp
\label{eightcross}
& u_{15} = {\chi^{+}\over 1 + \chi^{+}} \ , \qquad u_{26} \ = \ {\chi^{-}\over 1 + \chi^{-}}
\ ,
\qquad
u_{37} = {1\over 1 + \chi^{+}} \ , \qquad u_{48} \ = \ {1\over 1 + \chi^{-}}\ ,
\\
& u_{i\, i+3} = 1\ ,  \ \  \qquad \qquad i=1, \ldots , 8\,.
\esp\eeq
For more details on these kinematics and how to express the positions of the cusps of the Wilson loop in terms of $\chi^\pm$ we refer to Refs.~\cite{Brandhuber:2009da,Alday:2009yn}, as well as to Appendix~\ref{app:kinematics}.

Let us conclude this section by summarizing the properties of the remainder function in these special kinematics.
The invariance of the Wilson loop under a cyclic permutation of the edges implies that the remainder function $R_{8,WL}^{(2)}$, as a function of the twelve conformal ratios, must be symmetric under a simultaneous cyclic permutation or reversal of the first eight and the last four cross ratios in Eq.~(\ref{eq:us8})~\cite{Anastasiou:2009kna}. It is easy to check that in the $\chi^\pm$ kinematics this implies the invariance of the remainder function under a reversal and/or inversion of $\chi^+$ and $\chi^-$, \emph{i.e.},
\beq
\label{eq:R8sym}
R_{8,WL}^{(2)}(\chi^+,\chi^-) = R_{8,WL}^{(2)}(\chi^-,\chi^+) = R_{8,WL}^{(2)}(1/\chi^+,1/\chi^-) = R_{8,WL}^{(2)}(1/\chi^-,1/\chi^+)\,.
\eeq
Furthermore, the limits where the $\chi$ variables become large or small correspond to various soft and collinear limits~\cite{Alday:2009yn}. In general kinematics in the triple collinear limit, the octagon remainder function must reduce to the sum of two hexagon remainder 
functions~\cite{Anastasiou:2009kna}. In that limit and in the kinematics of \Eqn{eightcross}, we obtain that 
the eight-point remainder function must reduce to twice the remainder 
function for a regular hexagon~\cite{Brandhuber:2009da},
\beq
\label{eq:R82R6}
R_{8,WL}^{(2)} \rightarrow 2\,R_{6,\textrm{reg}}^{(2)} = -{\pi^4\over 18}\,.
\eeq

\subsection{The quasi-multi-Regge kinematics of four-of-a-kind along the ladder}

In Ref.~\cite{DelDuca:2008jg}, it was noted that the calculation of a six-point MHV amplitude (and thus of
a six-edged Wilson loop) would be exact if performed in multi-Regge kinematics which do not modify
the functional dependence of the cross ratios on the kinematic invariants. The simplest of such kinematics
are the quasi-multi-Regge kinematics (QMRK) of a pair along the ladder~\cite{Fadin:1989kf,DelDuca:1995ki}. 
This fact was used in
Refs.~\cite{DelDuca:2009au,DelDuca:2010zg} to compute analytically the two-loop six-edged Wilson loop.
In fact, in Ref.~\cite{DelDuca:2009au} it was shown that in the
QMRK of a cluster of $n-4$ particles along the ladder, an $L$-loop $n$-edged Wilson loop  is \emph{Regge exact},
{\it i.e.}, its analytic form is the same as in arbitrary kinematics\footnote{Regge exactness was noted
firstly in the simplest instance of the Regge limit of a four-edged Wilson loop~\cite{Drummond:2007aua}.
In fact, the Regge limit can be seen as the limiting case of the QMRK of a cluster
of zero particles along the ladder. However, in the case of a four-edged Wilson loop there are no cross
ratios and no remainder function, so the Regge exactness pertains only to the iterative part of \Eqn{eq:wl2wi}.}.

For an eight-edged Wilson loop, the appropriate kinematics are the QMRK
of four particles along the ladder~\cite{claude}. In the physical region, defining 1 and 2 as the incoming gluons,
with momenta $p_2=(p_2^+/2,0,0,p_2^+/2)$ and $p_1=(p_1^-/2,0,0,-p_1^-/2)$,
and $3,\ldots,8$ as the outgoing gluons, the kinematics are set by taking the outgoing gluons strongly
ordered in rapidity, except for a cluster of four along the ladder. The ordering is,
\begin{equation}
y_3 \gg y_4 \simeq y_5 \simeq y_6 \simeq y_7 \gg y_8\,,\qquad |p_{3\perp}| \simeq |p_{4\perp}|
\simeq |p_{5\perp}| \simeq|p_{6\perp}| \simeq|p_{7\perp}| \simeq|p_{8\perp}|\,
,\label{qmrk8trec}
\end{equation}
where the particle momentum $p$ is parametrised in terms of the rapidity $y$ and the azimuthal angle $\phi$,
$p=(|p_{\perp}|\cosh y, |p_{\perp}|\cos\phi, |p_{\perp}|\sin\phi,|p_{\perp}|\sinh y)$.
We shall work in the Euclidean region, where the Wilson loop is real.
In that case, the Mandelstam invariants can be taken as all negative, and in the QMRK of four-of-a-kind along the ladder
they are ordered as follows,
\bea
-s_{12} \gg -s_{1234}, -s_{3456}, -s_{123}, -s_{345}, -s_{678}, -s_{812}, -s_{34}, -s_{78} \gg  
&& \nn \\ &&  \hspace*{-105mm}
\gg -s_{2345}, -s_{4567}, -s_{234}, -s_{456}, -s_{567}, -s_{781},
-s_{23}, -s_{45},  -s_{56}, -s_{67}, -s_{81}\, .\label{eq:qmrk8quadceucl}
\eea
Introducing a parameter $\lambda \ll 1$, the hierarchy above is equivalent to the rescaling
\bea
&& \{ s_{1234}, s_{3456}, s_{123}, s_{345}, s_{678}, s_{812}, s_{34}, s_{78} \} = \ord(\lambda)\,, \nn\\
&& \{ s_{2345}, s_{4567}, s_{234}, s_{456}, s_{567}, s_{781}, s_{23}, s_{45}, s_{56}, s_{67}, s_{81} \} = \ord(\lambda^2)\,.
\eea
In this case, it is easy to verify that all the twelve cross ratios (\ref{eq:8cross}) are $\ord(1)$.
Because the dependence of the remainder function $R^{(2)}_{8,WL}$ on the twelve cross ratios is left
invariant in going from the exact kinematics to the QMRK of \Eqn{qmrk8trec}, these are the candidate simplest
kinematics by which to determine $R^{(2)}_{8,WL}$.


\section{The octagon remainder function in $\chi$ kinematics}
\label{sec:R8}
In this section we present our computation of the eight-edged remainder function in $\chi$ kinematics, which was done following the recipe introduced in Refs.~\cite{DelDuca:2009au, DelDuca:2010zg}. We start from the parametric representations of the Wilson loop diagrams given in Ref.~\cite{Anastasiou:2009kna} and derive appropriate Mellin--Barnes (MB) representations for all of them.
In multi-loop calculations it is sometimes difficult
to find an optimal choice for the MB representation. However, in our case the
MB representations are introduced in a straightforward way using the basic formula,
\begin{equation}
{1\over (A+B)^\lambda} =
{1\over \Gamma(\lambda)}\,\int_{-i\infty}^{+i\infty}{\rd z\over 2\pi i}\,
\Gamma(-z)\,\Gamma(\lambda+z)\,{A^{z}\over B^{\lambda+z}}\, ,
\label{MB}
\end{equation}
where the contour is chosen such as to separate the poles in $\Gamma(\ldots-z)$ from
the poles in $\Gamma(\ldots+z)$.
Note that in our case $\lambda$ equals an integer plus an off-set corresponding
to the dimensional regulator $\eps$. In order to resolve the singularity structures in
$\epsilon$, we apply the strategy based on the MB representation and given in
Refs.~\cite{Smirnov:1999gc,Tausk:1999vh,Smirnov:2004ym,Smirnov:2006ry}. To this effect,
we apply the codes {\tt MB}~\cite{Czakon:2005rk} and {\tt MBresolve}~\cite{Smirnov:2009up}
and obtain a set of MB integrals which can safely be expanded in $\eps$ under the integration sign. After applying these codes, all the integration contours are straight vertical lines. At the end of this procedure, the most complicated integral is expressed as a tenfold MB integral, which is dependent on ratios of Mandelstam invariants.

We then simplify the computation by exploiting the Regge exactness of the Wilson loop~\cite{DelDuca:2009au} and extract the leading quasi-multi-Regge behaviour by applying {\tt MBasymptotics} \cite{CzakonMBA}. Finally, we apply {\tt barnesroutines} \cite{Kosower} to perform
integrations that can be done by corollaries of Barnes lemmas. We arrive at a representation in terms of at most fivefold integrals depending explicitly on the cross ratios only\footnote{However, the coefficients of the integrals depend on logarithms of Mandelstam invariants.}. We checked numerically that the sum of the
MB integrals in the QMRK equals the sum of all the original parametric
integrals, the latter being evaluated numerically
using {\tt FIESTA}~\cite{Smirnov:2008py, Smirnov:2009pb}, as well as comparing directly to results obtained by the numerical code of Ref.~\cite{Anastasiou:2009kna}. It is worth noting that, although the individual integrals have undergone a huge simplification, due to the Regge exactness of the Wilson loop the representation of $w_8^{(2)}$ obtained in this way is valid in arbitrary kinematics.

The integrals we obtained can be simplified further by introducing the $\chi^\pm$ variables via Eq.~(\ref{eightcross}). Since most of the cross ratios become one in this limit, many of the MB integrals can be done in closed form using (corollaries of) Barnes lemmas, which, after some additional massaging, leaves us with at most twofold integrals to compute. All the integrals can now be computed by closing the contours at infinity and summing up the residues in the poles of the $\Gamma$ functions. The sums we obtain are nested harmonic sums~\cite{Moch:2001zr} that sum up to (multiple) polylogarithms, a task that can easily be performed using the {\tt FORM} code {\tt XSummer}~\cite{Moch:2005uc}\footnote{In intermediate steps, some of the integrals also get contributions from multiple binomial sums~\cite{Jegerlehner:2002em, Kalmykov:2007dk}. All of these terms cancel however in the sum over all contributions.}. Combining all the terms, and after a final massaging, we arrive at a very simple expression for the octagon remainder function,
\beq
\label{eq:R8chi}
R_{8,WL}^{(2)}(\chi^+,\chi^-) = -{\pi^4\over18} -{1\over 2}\ln\left(1+\chi^+\right)\ln\left(1+{1\over\chi^+}\right)\ln\left(1+\chi^-\right)\ln\left(1+{1\over\chi^-}\right)\,.
\eeq
Eq.~(\ref{eq:R8chi}) is the main result of this paper. We checked its correctness by comparing to various points obtained by the numerical code of Ref.~\cite{Brandhuber:2009da}\footnote{We are grateful to Paul Heslop and Valya Khoze for providing us with this check.}. Note that in
Eq.~(\ref{eq:R8chi}) the symmetry properties (\ref{eq:R8sym}) of the remainder function are manifest. Furthermore, in the limit where one of the $\chi$ variables becomes large or small, Eq.~(\ref{eq:R8chi}) immediately reduces to $-{\pi^4\over 18}$, in agreement with Eq.~(\ref{eq:R82R6}).
Finally, we can extract from Eq.~(\ref{eq:R8chi}) the value of the regular octagon, which corresponds to $\chi^\pm=1$,
\beq
\label{eq:R8chi11}
R_{8,WL}^{(2)}(1,1) = -{\pi^4\over18} -{1\over 2}\ln^42 \simeq -5.52703\ldots\,,
\eeq
in a very good agreement with the numerical value quoted in Ref.~\cite{Brandhuber:2009da}.

\section{Comparison to the strong coupling result}
\label{sec:strong}

\begin{center}
\begin{figure}[!t]
\begin{center}
\begin{minipage}[c]{7.5cm}
\includegraphics[scale=0.85]{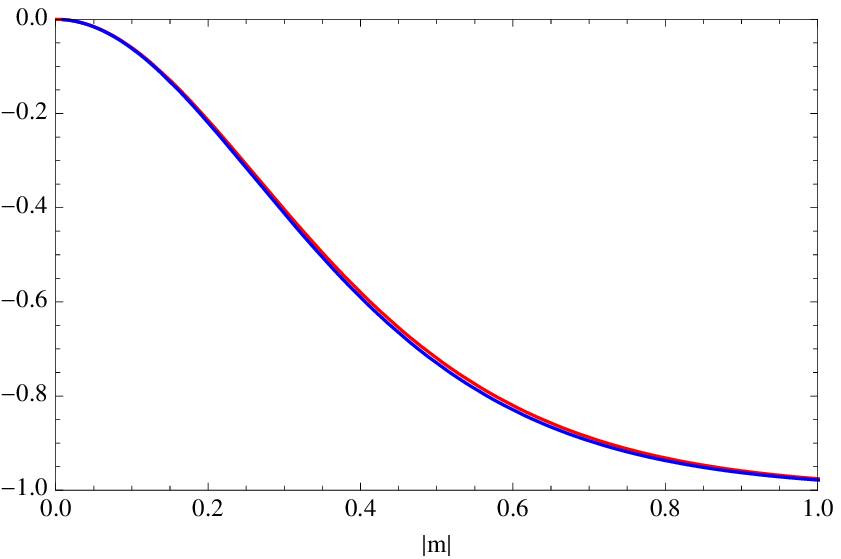}
\end{minipage}
\begin{minipage}[c]{7.5cm}
\includegraphics[scale=0.84]{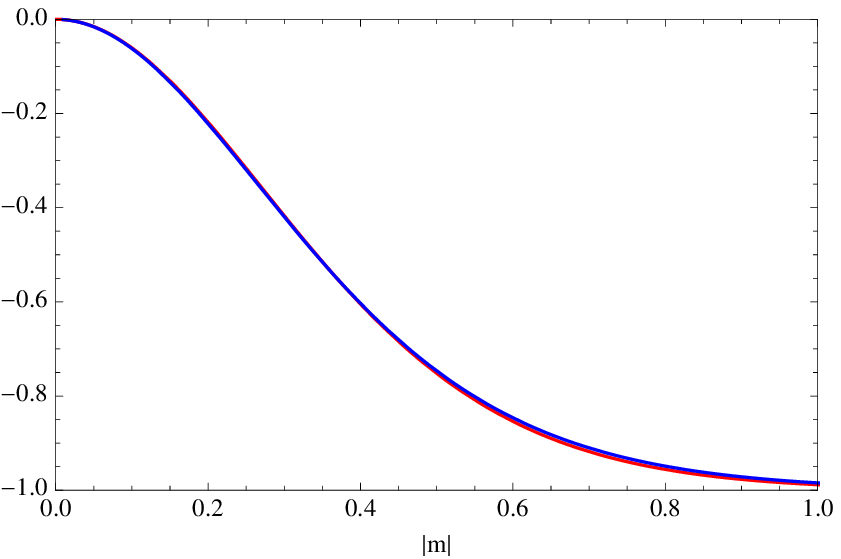}
\end{minipage}
\caption{\label{fig:diff_m}Plot for $\overline{R}_{8}^{(2)}$ (red) and $\overline{R}_{8}^\textrm{strong}$ (blue) as a function of $|m|$ for $\phi = 0$ (left) and $\phi = \pi/4$ (right). Note that the two curves basically overlap and that the numerical difference between them is too small to be appreciated by eye.}
\end{center}
\end{figure}
\end{center}

In Ref.~\cite{Alday:2009yn} the strong coupling octagon remainder function was given in terms of a one-dimensional integral,
\beq\bsp\label{eq:R8strong}
R_{8,WL}^\textrm{strong} =&\,-{1\over2}\ln\left(1+\chi^-\right)\ln\left(1+{1\over\chi^+}\right) + {7\pi\over6} \\
&\,+ \int_{-\infty}^{+\infty}\rd t\,{|m|\,\sinh t\over \tanh(2t+2i\phi)}\,\ln\left(1+e^{-2\pi |m| \cosh t}\right)\,,
\esp\eeq
where $m = |m|e^{i\phi}$ is a complex variable related to $\chi^{\pm}$ via
\beq
\chi^+ = e^{2\pi \textrm{Im}\, m} {\rm ~~and~~} \chi^- = e^{-2\pi \textrm{Re}\, m}\,.
\eeq
Eq.~(\ref{eq:R8strong}) is valid in the first quadrant of the complex $m$-plane, $0<\phi<{\pi\over 2}$, and is extended over the whole complex plane by analytic continuation. Note that Eq.~(\ref{eq:R8strong}) is invariant under $\phi\rightarrow\phi+{\pi\over 2}$, reflecting the invariance of the remainder function under exchange and inversion of the $\chi$ variables. In the collinear limit, Eq.~(\ref{eq:R8strong}) reduces to twice the remainder function of the regular hexagon,
\beq
R_{8,WL}^\textrm{strong} \rightarrow 2R_{6,\textrm{reg}}^\textrm{strong} = {7\pi\over 6}\,.
\eeq
Furthermore, the value of the regular octagon is also known analytically,
\beq
R_{8,\textrm{reg}}^\textrm{strong} = {5\pi\over 4} - {1\over 2}\ln^22\,.
\eeq

In Ref.~\cite{Brandhuber:2009da} the following rescaled remainder function was introduced, both at weak and at strong coupling,
\beq
\overline{R}_{8}^i = {R_{8,WL}^i - R^i_{8,\textrm{reg}} \over R^i_{8,\textrm{reg}} - 2R^i_{6,\textrm{reg}}}\,,
\eeq
where $i$ refers either to the strong or the weak coupling answer. It was observed that within numerical errors this rescaled remainder function is equal at weak and at strong coupling,
\beq
\overline{R}_{8}^\textrm{strong} \simeq \overline{R}_{8}^{(2)}\,,
\eeq
and it was conjectured that such a universality might extend even beyond the case of the octagon and/or the special kinematics under consideration.

\begin{center}
\begin{figure}[!t]
\begin{center}
\begin{minipage}[c]{7.5cm}
\includegraphics[scale=0.85]{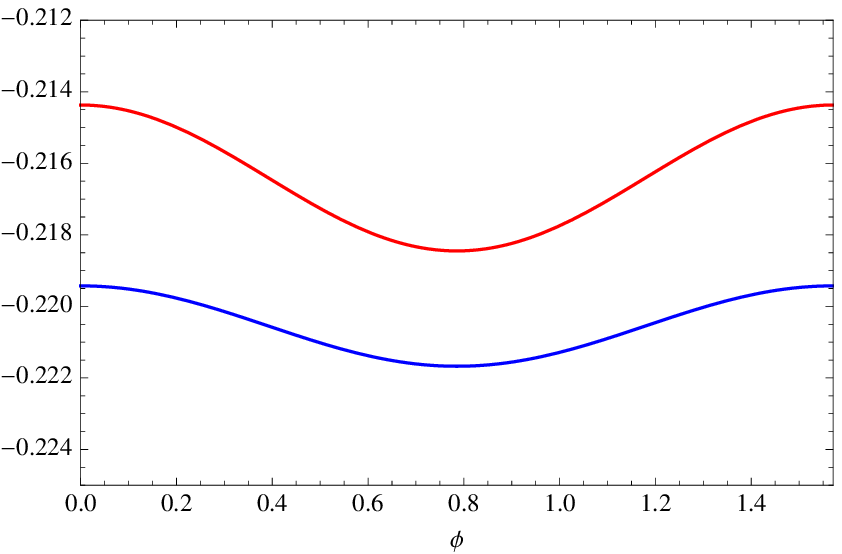}
\end{minipage}
\begin{minipage}[c]{7.5cm}
\includegraphics[scale=0.84]{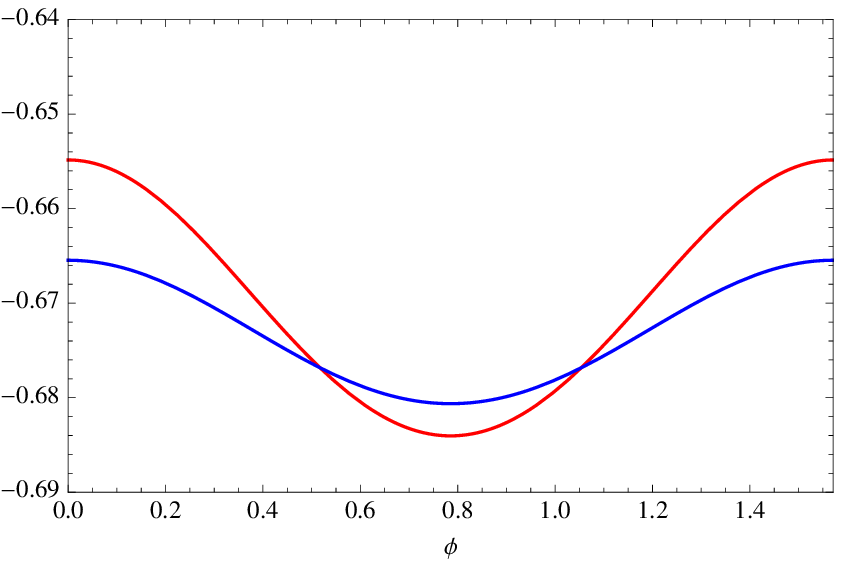}
\end{minipage}
\caption{\label{fig:diff_phi}Plot for $\overline{R}_{8}^{(2)}$ (red) and $\overline{R}_{8}^\textrm{strong}$ (blue) as a function of $\phi$ for $|m| = 0.2$ (left) and $|m| = 0.45$ (right).}
\end{center}
\end{figure}
\end{center}

Since we are now in possession of an analytic expression for the weak coupling result, we can check this conjecture to a much higher accuracy. We find that, similar to the case of the hexagon remainder function, the two functions are indeed very close over a wide range of values, but they differ substantially not only in magnitude, but also in shape (See Fig.~\ref{fig:diff_m} and~\ref{fig:diff_phi}).


\section{Conclusion}
\label{sec:conclusion}
In this paper we have presented the first analytic computation of the two-loop remainder function for an eight-edged Wilson loop in $\begin{cal}N\end{cal}=4$ SYM, in the kinematic set-up of Ref.~\cite{Alday:2009yn}. The result is characterised by a remarkably simple form, a constant plus a product of four logarithms. In fact it corresponds to the simplest function of uniform transcendentality four consistent with the constraints coming from cyclic invariance and collinear limits, Eqs.~(\ref{eq:R8sym}) and (\ref{eq:R82R6}). Hence its functional form is simpler than the strong coupling result of Ref.~\cite{Alday:2009yn}. This is to be contrasted with the case of the hexagon remainder function, where the strong coupling result~\cite{Alday:2009dv} is given by a rather short expression in terms of simple functions, whereas the weak coupling result~\cite{DelDuca:2009au,DelDuca:2010zg} is expressed as a complicated combination of polylogarithms of weight four. 

In order to test the conjecture of a universality of the remainder function in the strong  and weak coupling limits,
in Section~\ref{sec:strong} we confronted our result to the strong coupling result. In contrast with the numerical observation of
Ref.~\cite{Brandhuber:2009da}, we do not find a matching between the strong and weak coupling results, because both functions differ not only in magnitude but also in shape. Thus we conclude that a universality of the type suggested in Ref.~\cite{Brandhuber:2009da} is ruled out. 

In the perspective of studying further potential relations between the strong- and weak-coupling remainder functions,
it would be interesting to compare our result to the recently introduced OPE approach to polygonal Wilson loops~\cite{Alday:2010ku}.

\section*{Acknowledgements}

We thank Fernando Alday, Paul Heslop, Valya Khoze and Amit Sever for useful discussions.
This work was partly supported by RFBR, grant 08-02-01451, and
by the EC Marie-Curie Research Training Network ``Tools and Precision
Calculations for Physics Discoveries at Colliders'' under contract MRTN-CT-2006-035505.


\newpage
\appendix

\section{Eight-point kinematics}
\label{app:kinematics}

The diagrams that enter the computation of the two-loop eight-edged Wilson loop have been given explicitly in
Ref.~\cite{Anastasiou:2009kna}. In terms of those diagrams, we write the Wilson loop as,
\bea
w_8^{(2)} &=& \cC \left[
f_H(p_1,p_2,p_3; 0, p_4+p_5+p_6+p_7+p_8, 0) + f_H(p_1,p_2,p_4; p_3, p_5+p_6+p_7+p_8, 0) \right. \nn\\
&& + f_H(p_1,p_2,p_5; p_3+p_4,p_6+p_7+p_8, 0) + f_H(p_1,p_2,p_6; p_3+p_4+p_5,p_7+p_8, 0) \nn\\
&& + f_H(p_1,p_2,p_7; p_3+p_4+p_5+p_6,p_8, 0) + f_H(p_1,p_3,p_5; p_4,p_6+p_7+p_8,p_2) \nn\\
&& + f_H(p_1,p_3,p_6; p_4+p_5,p_7+p_8,p_2) \nn\\
&& + f_C(p_1,p_2,p_3; 0, p_4+p_5+p_6+p_7+p_8, 0) + f_C(p_1,p_2,p_4; p_3, p_5+p_6+p_7+p_8, 0) \nn\\
&& + f_C(p_1,p_2,p_5; p_3+p_4,p_6+p_7+p_8, 0) + f_C(p_1,p_2,p_6; p_3+p_4+p_5, p_7+p_8,0) \nn\\
&& + f_C(p_1,p_2,p_7; p_3+p_4+p_5+p_6,p_8,0) +  f_C(p_1,p_2,p_8; p_3+p_4+p_5+p_6+p_7,0,0) \nn\\
&& + f_C(p_1,p_3,p_4; 0, p_5+p_6+p_7+p_8, p_2) + f_C(p_1,p_3,p_5; p_4, p_6+p_7+p_8, p_2) \nn\\
&& + f_C(p_1,p_3,p_6; p_4+p_5, p_7+p_8, p_2) + f_C(p_1,p_3,p_7; p_4+p_5+p_6, p_8, p_2) \nn\\
&& + f_C(p_1,p_3,p_8; p_4+p_5+p_6+p_7, 0, p_2) + f_C(p_1,p_4,p_5; 0, p_6+p_7+p_8, p_2+p_3) \nn\\
&& + f_C(p_1,p_4,p_6; p_5, p_7+p_8, p_2+p_3) + f_C(p_1,p_4,p_7; p_5+p_6, p_8, p_2+p_3) \nn\\
&& + f_C(p_1,p_4,p_8; p_5+p_6+p_7, 0, p_2+p_3) + f_C(p_1,p_5,p_6; 0, p_7+p_8, p_2+p_3+p_4) \nn\\
&& + f_C(p_1,p_5,p_7; p_6, p_8, p_2+p_3+p_4) + f_C(p_1,p_5,p_8; p_6+p_7, 0, p_2+p_3+p_4) \nn\\
&& + f_C(p_1,p_6,p_7; 0, p_8, p_2+p_3+p_4+p_5) + f_C(p_1,p_6,p_8; p_7, 0, p_2+p_3+p_4+p_5) \nn\\
&& +  f_C(p_1,p_7,p_8; 0, 0, p_2+p_3+p_4+p_5+p_6) \nn\\
&& + f_X(p_1,p_2; p_3+p_4+p_5+p_6+p_7+p_8, 0) + f_Y(p_1,p_2; p_3+p_4+p_5+p_6+p_7+p_8, 0) \nn\\
&& + f_Y(p_2,p_1; 0, p_3+p_4+p_5+p_6+p_7+p_8) + f_X(p_1,p_3; p_4+p_5+p_6+p_7+p_8, p_2) \nn\\
&& + f_Y(p_1,p_3; p_4+p_5+p_6+p_7+p_8, p_2) + f_Y(p_3,p_1; p_2, p_4+p_5+p_6+p_7+p_8) \nn\\
&& + f_X(p_1,p_4; p_5+p_6+p_7+p_8, p_2+p_3) +  f_Y(p_1,p_4; p_5+p_6+p_7+p_8, p_2+p_3) \nn\\
&& + f_Y(p_4,p_1; p_2+p_3, p_5+p_6+p_7+p_8) +  (1/2) f_X(p_1,p_5; p_6+p_7+p_8, p_2+p_3+p_4) \nn\\
&& + f_Y(p_1,p_5; p_6+p_7+p_8, p_2+p_3+p_4) \nn\\
&& + (-1/2) f_P(p_1,p_3; p_4+p_5+p_6+p_7+p_8, p_2)\, f_P(p_2,p_4; p_1+p_5+p_6+p_7+p_8, p_3) \nn\\
&& + (-1/2) f_P(p_1,p_3; p_4+p_5+p_6+p_7+p_8, p_2)\, f_P(p_2,p_5; p_1+p_6+p_7+p_8, p_3+p_4) \nn\\
&& + (-1/2) f_P(p_1,p_3; p_4+p_5+p_6+p_7+p_8, p_2)\, f_P(p_2,p_6; p_1+p_7+p_8, p_3+p_4+p_5) \nn\\
&& + (-1/2) f_P(p_1,p_3; p_4+p_5+p_6+p_7+p_8, p_2)\, f_P(p_2,p_7; p_1+p_8, p_3+p_4+p_5+p_6) \nn\\
&& + (-1/2) f_P(p_1,p_4; p_5+p_6+p_7+p_8, p_2+p_3)\, f_P(p_2,p_5; p_1+p_6+p_7+p_8, p_3+p_4) \nn\\
&& + (-1/2) f_P(p_1,p_4; p_5+p_6+p_7+p_8, p_2+p_3)\, f_P(p_2,p_6; p_1+p_7+p_8, p_3+p_4+p_5) \nn\\
&& + (-1/2) f_P(p_1,p_4; p_5+p_6+p_7+p_8, p_2+p_3)\, f_P(p_2,p_7; p_1+p_8, p_3+p_4+p_5+p_6) \nn\\
&& + (-1/2) f_P(p_1,p_4; p_5+p_6+p_7+p_8, p_2+p_3)\, f_P(p_3,p_7; p_1+p_2+p_8, p_4+p_5+p_6) \nn\\
&& + (-1/4) f_P(p_1,p_5; p_6+p_7+p_8, p_2+p_3+p_4)\, f_P(p_2,p_6; p_1+p_7+p_8, p_3+p_4+p_5) \nn\\
&& + (-1/8) f_P(p_1,p_5; p_6+p_7+p_8, p_2+p_3+p_4)\, f_P(p_3,p_7; p_1+p_2+p_8, p_4+p_5+p_6) \nn\\
&&  \left. +\ {\rm cyclic\ permutations\ of}\ (p_1,p_2,p_3, p_4, p_5,p_6,p_7,p_8) \right]\,,
\eea
where, in the terminology of  Ref.~\cite{Anastasiou:2009kna}, $f_H$ stands for a hard diagram, $f_C$ for a curtain
diagram, $f_X$ for a cross diagram, $f_Y$ for a Y diagram plus half a self-energy diagram,
$f_P$ for a factorised cross diagram. Furthermore,
\beq
\cC = 2 a^2 \mu^{4\eps} \left[ \Gamma(1+\eps) e^{\gamma\eps}\right]^2\,,
\eeq
and the scale $\mu^2$ is given in terms of the Wilson loop scale, $\mu_{WL}^2 = \pi e^\gamma \mu^2$.

The Wilson loop $w_8^{(2)}$ is expressed in terms of the kinematic invariants, $s_{12}$, $s_{23}$, $s_{34}$,
$s_{45}$, $s_{56}$, $s_{67}$, $s_{78}$, $s_{81}$, $s_{123}$, $s_{234}$, $s_{345}$, $s_{456}$, $s_{567}$,
$s_{678}$, $s_{781}$, $s_{812}$, $s_{1234}$, $s_{2345}$, $s_{3456}$, $s_{4567}$,
all the other invariants being related to those above by the following relations,
\bea
s_{13} &=& s_{123} -s_{12} - s_{23}\,, \qquad s_{14} = s_{1234} - s_{123} - s_{234} + s_{23}\,, \nn \\
s_{15} &=& - s_{1234} - s_{2345} + s_{234} + s_{678}\,, \qquad s_{16} = s_{2345} - s_{678} - s_{781} + s_{78}\,, \nn\\
s_{17} &=& s_{781} -s_{78} - s_{81}\,, \qquad s_{24} = s_{234} -s_{23} - s_{34}\,, \nn\\
s_{25} &=& s_{2345} - s_{234} - s_{345} + s_{34}\,, \qquad s_{26} = - s_{2345} - s_{3456} + s_{345} + s_{781}\,, \nn\\
s_{27} &=& s_{3456} - s_{781} - s_{812} + s_{81}\,, \qquad s_{28} = s_{812} - s_{81} - s_{12}\,, \nn \\
s_{35} &=& s_{345}  -s_{34} - s_{45}\,, \qquad s_{36} = s_{3456} - s_{345} - s_{456} + s_{45}\,, \nn \\
s_{37} &=& - s_{3456} - s_{4567} + s_{456}  + s_{812}\,, \qquad s_{38} = s_{4567} - s_{812} - s_{123} + s_{12}\,, \nn\\
s_{46} &=& s_{456} - s_{45} - s_{56}\,, \qquad s_{47} = s_{4567} - s_{456}  -s_{567}  + s_{56}\,, \nn\\
s_{48} &=& - s_{4567} - s_{1234} + s_{567}  + s_{123}\,, \qquad s_{57} = s_{567} - s_{56} - s_{67}\,, \nn\\
s_{58} &=& s_{1234} - s_{567}  -s_{678}  + s_{67}\,, \qquad s_{68} = s_{678} - s_{67} - s_{78}\,.
\eea
Using these expressions, we can form twelve independent conformal cross ratios,
\bea
&& u_{14} = \frac{x_{15}^2 x_{24}^2}{x_{14}^2 x_{25}^2} =
\frac{s_{1234}s_{23}}{s_{123}s_{234}}\,, \quad
u_{25} = \frac{x_{26}^2 x_{35}^2}{x_{25}^2 x_{36}^2} =
\frac{s_{2345}s_{34}}{s_{234}s_{345}}\,, \quad
u_{36} = \frac{x_{37}^2 x_{46}^2}{x_{36}^2 x_{47}^2} =
\frac{s_{3456}s_{45}}{s_{345}s_{456}}\,, \nn\\
&& u_{47} = \frac{x_{48}^2 x_{57}^2}{x_{47}^2 x_{58}^2} =
\frac{s_{4567}s_{56}}{s_{456}s_{567}}\,, \quad
u_{58} = \frac{x_{51}^2 x_{68}^2}{x_{58}^2 x_{61}^2} =
\frac{s_{1234}s_{67}}{s_{567}s_{678}}\,, \quad
u_{61} = \frac{x_{62}^2 x_{71}^2}{x_{61}^2 x_{72}^2} =
\frac{s_{2345}s_{78}}{s_{678}s_{781}}\,, \label{eq:8cross}\\
&& u_{72} = \frac{x_{73}^2 x_{82}^2}{x_{72}^2 x_{83}^2} =
\frac{s_{3456}s_{81}}{s_{781}s_{812}}\,, \quad
u_{83} = \frac{x_{84}^2 x_{13}^2}{x_{83}^2 x_{14}^2} =
\frac{s_{4567}s_{12}}{s_{812}s_{123}}\,, \quad
u_{15} = \frac{x_{16}^2 x_{25}^2}{x_{15}^2 x_{26}^2} =
\frac{s_{678}s_{234}}{s_{1234}s_{2345}}\,, \nn\\
&& u_{26} = \frac{x_{27}^2 x_{36}^2}{x_{26}^2 x_{37}^2} =
\frac{s_{781}s_{345}}{s_{2345}s_{3456}}\,, \quad
u_{37} = \frac{x_{38}^2 x_{47}^2}{x_{37}^2 x_{48}^2} =
\frac{s_{812}s_{456}}{s_{3456}s_{4567}}\,, \quad
u_{48} = \frac{x_{41}^2 x_{58}^2}{x_{48}^2 x_{51}^2} =
\frac{s_{123}s_{567}}{s_{4567}s_{1234}}\,, \nn
\eea
where we have used \Eqn{eq:dist} to set $x_{i,j}^2 = s_{i,i+1,\ldots,j-1}$, with $j > i$.

In Ref.~\cite{Brandhuber:2009da}, three representations of the set-up of Ref.~\cite{Alday:2009yn}
in terms of the Wilson-loop cusp coordinates are given.
We choose the one for which the cusp coordinates are
\begin{align}
x_1 &= (1/2, 1/2, -1)\, , \qquad
  x_2 = \left({\chi^+\over 2 + 2 \chi^+}, {\chi^+\over 2 + 2 \chi^+}, -1\right)\, ,
  \nonumber\\
x_3 &= \left({1 + (2 + \chi^-) \chi^+\over
   2 (1 + \chi^- + \chi^- \chi^+)}, {-1 + \chi^- \chi^+\over
   2 (1 + \chi^- + \chi^- \chi^+)}, {-(1 + \chi^-) (1 + \chi^+)\over
    1 + \chi^- + \chi^- \chi^+}\right)\, , \nonumber\\
x_4 &= \left({1\over 2 + 2 \chi^-}, {-1\over 2 (1 + \chi^-)}, -1\right)\, ,\qquad
x_5 = (1/2, -1/2, -1),\nonumber\\
x_6 &= (-1/2, -1/2, 0)\, ,\qquad
x_7 = (0, 0, 0)\, ,\qquad
x_8 = (-1/2, 1/2, 0)\, .
\label{8ptkinematics}
\end{align}
Then the Mandelstam invariants take the values
\begin{align}
x_{13}^2 &= s_{12} = - \frac{1}{1+\chi^-(1+\chi^+)}\ ,\qquad x_{14}^2 = s_{123} = - \frac{1}{1+\chi^-}\ ,\nn\\
x_{15}^2 &= s_{1234} = -1\ , \qquad x_{16}^2 = s_{678} = -1\ ,\qquad x_{17}^2 = s_{78} = -1\ ,\nn\\
x_{24}^2 &= s_{23} = - \frac{\chi^+}{(1+\chi^-)(1+\chi^+)}\ ,\qquad
x_{25}^2 = s_{234} = - \frac{\chi^+}{1+\chi^+}\ ,\nn\\
x_{26}^2 &= s_{2345} = -1\ ,\qquad x_{27}^2 = s_{781} = -1\ ,\qquad x_{28}^2 = s_{81} = - \frac{1}{1+\chi^+}\ ,\nn\\
x_{35}^2 &= s_{34} = - \frac{\chi^+\chi^-}{1+\chi^-(1+\chi^+)}\ ,\qquad
x_{36}^2 = s_{345} = - \frac{\chi^-(1+\chi^+)}{1+\chi^-(1+\chi^+)}\ ,\nn\\
x_{37}^2 &= s_{3456} = - 1 - \frac{\chi^+}{1+\chi^-(1+\chi^+)}\ ,\qquad
x_{38}^2 = s_{812} = - \frac{1+\chi^-}{1+\chi^-(1+\chi^+)}\ ,\nn\\
x_{46}^2 &= s_{45} = - \frac{\chi^-}{1+\chi^-}\ ,\qquad x_{47}^2 = s_{456} = -1\ ,\qquad x_{48}^2 = s_{4567} = -1\ ,\nn\\
x_{57}^2 &= s_{56} = - 1\ ,\qquad x_{58}^2 = s_{567} = - 1\ ,\qquad x_{68}^2 = s_{67} = - 1\ .
\label{8ptmandelstam}
\end{align}
Using these expression, we can immediately construct the cross ratios given in Eq.~(\ref{eightcross}).

\end{document}